\documentclass[showpacs,aps,superscriptaddress,prb,twocolumn,amsmath]
{revtex4}
\usepackage{txfonts}
\usepackage{color}
\usepackage{amssymb}
\usepackage{graphicx}% Include figure files
\usepackage{dcolumn}% Align table columns on decimal point
\usepackage{bm}% bold math
\usepackage{hyperref}% add hypertext capabilities
\begin{document}

\preprint{APS/123-QED}

\title{Magnetic Exchange Coupling and  Anisotropy of 3d Transition-Metal Nanowire on the Surface of Graphyne Sheet}
%\title{\emph{sp}-hybridized Carbon-mediated Transition-Metal chain based on
%graphyne:magnetic property.}
%====================================================================
\author{Junjie He}
\affiliation{Department of Physics, Xiangtan University, Xiangtan
411105, China}
\author{Pan Zhou}
\affiliation{Department of Physics, Xiangtan University, Xiangtan
411105, China}
\author{N. Jiao}
\affiliation{Department of Physics, Xiangtan University, Xiangtan
411105, China}
\author{S. Y. Ma}
\affiliation{Department of Physics, Xiangtan University, Xiangtan
411105, China}
\author{K. W. Zhang}
\affiliation{Department of Physics, Xiangtan University, Xiangtan 411105, China}
\author{R. Z. Wang}
\affiliation{College of Materials Science and Engineering, Beijing University of Technology, Beijing100124, China}
\author{L. Z. Sun}
\email{lzsun@xtu.edu.cn} \affiliation{Department of Physics,
Xiangtan University, Xiangtan 411105, China}

\date{\today}% It is always \today, today,

\begin{abstract}
\indent Using density functional theory plus  Hubbard-U (DFT+U) approach, we find that quasi one-dementation(1D) 3d transition metal(TM) zigzag nanowire can be constructed by TM adsorbed on the surface of graphyne sheet. The results show that the TM exchange coupling of the zigzag nanowire mediated by \emph{sp} hybridized carbon atoms gives rise to long range ferromagnetic order except for Cr with anti-ferromagnetic order. The magnetic exchange interaction of TM chains follows like-Zener's $p_z$-d exchange mechanism: the coexistence of out-of plane $p_z$-d and in-plane $p_{x-y}$-d exchange. Finally, by including spin-orbit interactions within spin-DFT, we calculate the magnetic anisotropy energy of the TM chain on graphyne. We find that the Fe and Co chains show considerable magnetic anisotropy energy (MAE) and orbital magnetic moment. The easy axis of V, Cr, Mn and Fe chains is perpendicular to the surface, whereas the easy axis of Co lies in the surface. Moreover, only V chain shows relatively larger in-plane anisotropy. Our results open a new route to realize the applications of graphyne in spintronics.\\
\end{abstract}
 \maketitle
%++++++++++++++++++++++++++++++++++++++++++++++++++++++++++++w
\section{Introduction}
\indent Since the synthesis of two-dimensional (2D) graphene\cite{1,2}, much research has been focused on it due to its unique properties and tremendous possible applications in nanoscale devices.\cite{3,4,5} Graphyne (GY), a 2D carbon allotrope with the same symmetry as graphene predicted to have a high possibility of synthesis\cite{6,7}, is made up of hexagonal carbon rings and acetylene linkages. Graphdiyne(GDY), belonging to the family of GY has been successfully grown on the surface of copper recently.\cite{8,9} Consequently, the preparation of GY can be expected because theoretical work has shown that it is energetically more stable than GDY. Previous work has shown that the surface adsorption of TM atom profoundly influences the electronic properties of graphene and induces spin polarization\cite{10,11,12,13,14,15} in the system, which makes it a vital candidate for spintronics\cite{16,17,18}. Likewise, in our previous work\cite{19}, the adsorption of single 3d TM atom on GY sheet will promote magnetization of the TM-GY system, which indicates that the TM-GY systems are potential candidates for spintronics. Moreover, recent studies based on DFT+U approach\cite{18}, auxiliary-field quantum Monte Carlo\cite{23}, and multireference quantum chemical methods\cite{24} have shown a weak binding for TM adatom on graphene. It is consistent with the transmission electron microscopy experiments\cite{25,26,27} that TM adatoms are fast surface diffuser on graphene. Porter et al.\cite{28} indicated that clustering of the TM adatoms is energetically favorable for TM on graphene sheet. Therefore, it is difficult to obtain stable magnetic order by TM absorbed on graphene. Different from the case of TM adsorbed on graphene, the existence of additional $\emph{p}_x$-$\emph{p}_y$$\pi$/$\pi^{\ast}$ state of the \emph{sp} hybridized carbon in GY makes it strongly bind to TM adatoms and easily trap the TM adatoms in the alkyne rings of GY.\cite{19} Compared with the cluster effect of TM adatoms on graphene, dispersed adsorption of the TM adatoms on GY is energetically favor. It is well known that the long-range characteristics of the indirectly exchange coupling between magnetic adatoms mediated by hosts material play a significant role in determining the magnetic order of TM on real materials. For example, between two magnetic impurity in graphene follows Ruderman-Kittel-Kasuya-Yoshida (RKKY) exchange coupling, which derived from the semi-metallic nature of graphene.\cite{20,21,22} To open the way for TM-GY applying in spintronics, it is essential to understand the magnetic alignment of TM adatoms adsorbed on the GY, especially their magnetic exchange coupling mediated by the \emph{sp} hybridized carbon atoms.\\
%+++++++++++++++++++++++++++++++++++++++++++++++++++++++++++++++++++++++++
\indent One dimensional (1D) TM monoatomic chain, whose magnetic properties are much different from those of bulk solid materials because of significant reduction in dimensionality, shows potential applications in spintronic devices, such as spin-dependent quantum transport properties, ballistic anomalous magnetoresistance,\cite{57} spin and magnetization tunneling\cite{58}. Moreover, much researches\cite{52,59,60} have shown that the 1D TM metallic chain can lead to large magnetic anisotropy. In present work, using density functional theory plus Hubbard-U (DFT+U) approach we find that the TM adatoms on GY can be constructed as TM chain indicating that the GY is a potential template for synthesizing TM nanowire. Such TM chain can behave as a metallic nanowire with spin-polarized currents or quantum channel bringing a new opportunity to construct integrated circuits using various combinations of connected monoatomic metallic wire based on GY. Moreover, such monoatomic TM chains can be created on the surface of GY by STM tip manipulation\cite{29,30,31,32,33,34}. In view of the strong intermediate of \emph{sp} hybridized carbon atom, it is easy to retain a long range magnetic order on the surface of GY. The results in present work show that the TM exchange coupling of the zigzag TM nanowire mediated by \emph{sp} hybridized carbon atoms gives rise to long range ferromagnetic order except for Cr with anti-ferromagnetic order. The magnetic exchange interaction of TM chains follows like-Zener's $p_z$-d exchange mechanism: the coexistence of out-of plane $p_z$-d and in-plane $p_{x-y}$-d exchange. Furthermore, we find that Fe and Co chains show considerable magnetic anisotropy energy (MAE) and orbital magnetic moment. The easy axis of the V, Cr, Mn and Fe chains is perpendicular to the surface, whereas the easy axis of Co lies in the surface. Moreover, only V chain shows relatively larger in-plane anisotropy. Our results open a new route to realize the applications of graphyne in spintronics.\\
%+++++++++++++++++++++++++++++++++++++++++++++++++++++++++++++++++++++++++
\indent The remainder of the paper is organized as follows. Section II describes our computational methods. The results and discussions are presented in Sec. III. In Sec. III (A) we discuss the adsorption configuration of TMs on the surface of GY and their stability. The electronic structures, magnetic properties and the magnetic exchange coupling mechanism of TM chain on the surface of GY are presented in Sec. III (B) . In Sec. III (C) we focus on the magnetic anisotropy of TM chain on the surface of GY. Section IV gives a brief summary of this work.\\
%+++++++++++++++++++++++++++++++++++++++++++++++++++++++++++++++++++++++++
\begin{figure*}
 \includegraphics[width=4.5in]{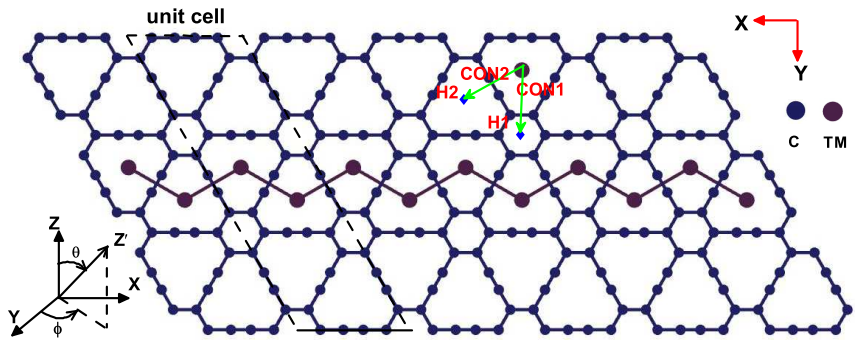}
 \caption{Schematic of the unit cell, TM-chain-GY, and TM dimer configuration CON1 and CON2. The sketch of the Cartesian and polar coordinate is used for the definition of the magnetization direction ($Z'$). Angles $\theta$ and $\phi$ represent the azimuthal and polar angles, respectively.}
 \label{fig1}
\end{figure*}
%+++++++++++++++++++++++++++++++++++++++++++++++++++++++++++++++++++++++++
\section{METHOD AND COMPUTATIONAL DETAILS}
\indent The unit cells of GY with TM adatom dimer and chain adopted in present work are shown in Fig.\ref{fig1}. A vacuum space of $13\AA$ perpendiculars to the GY plane is chosen to avoid the interactions between the neighboring images. The first-principles calculations are performed using the Vienna ab initio simulation package (VASP)\cite{35,36} within spin-polarized DFT.\cite{37,38} To account for the correlation energy of the strongly localized \emph{3d} orbital of TM, Hubbard U correction (DFT+U)\cite{39} is employed based on the generalized gradient approximation of Perdew-Burke-Ernzerhof (PBE).\cite{40} We choose the projector augmented wave (PAW) potentials\cite{41,42} to describe the electron-ionic core interaction. A plane-wave basis set with the kinetic energy cutoff of 400 eV is employed. The Brillouin zone (BZ) is sampled using 5$\times$3$\times$1 and 9$\times$5$\times$1 Gamma-centered Monkhorst-Pack grids for the calculations of relaxation and electronic structures, respectively. The criteria of energy and atom force convergence are set to $10^{-5}$ eV/unit cell and 0.01 eV/$\AA$, respectively. The dipole correction\cite{43} is considered to deal with the impact of variety of potential distribution introduced by TM adsorption. All calculation parameters in present work are systematically optimized. As for the calculations within DFT+U, the rotationally invariant DFT+U formalism proposed by Dudarev et al.\cite{44} is used, where only $U_{eff} =U-J$ value is meaningful instead of individual U and J values. We assign the value $U_{eff}$=4.84eV, 3.55eV, 5.01eV, 5.03eV and 6.31eV for V, Cr, Mn, Fe and Co adatom based on our earlier calculations\cite{19} with linear response method\cite{45}. The appropriateness of the uniform $U_{eff}$ value for single TM, TM dimer and TM chain is due to the localization of coupling effect of TM adatoms in TM dimer and TM chain, which will be discussed below. For the magnetic anisotropy calculations, the spin-orbit coupling (SOC) is included self-consistently. The MAE is calculated from the ground state energies for magnetization aligned in the most relevant directions. The total energies are converged to a precision of $10^{-7}$ eV.\\
%+++++++++++++++++++++++++++++++++++++++++++++++++++++++++++++++++++++++++
\indent To evaluate the stability of adsorption configuration, adsorption energy per TM atom E$_a$ is adopted which is defined as,
%++++++++++
\begin{equation}
E_{a}=\frac{1}{n}[({E_{GY}+nE_{TM})-E_{TM+GY}}]
\end{equation}
%++++++++++
where $E_{TM+GY}$ denotes spin-polarized total energy of TM adsorbed GY sheet, $E_{GY}$ is the total energy of isolated pure GY sheet, and $E_{TM}$ is the spin-polarized total energy of corresponding free TM atom.\\
%++++++++++++++++++++++++++++++++++++++x+++++++++++++++++++++++++++++++++++
\indent To analyze spin polarization of the systems, we define the spin polarization $P(E_{f})$ at the Fermi level as:
%++++++++++++
\begin{equation}
P(E_{f})=\frac{D(E_{f\uparrow})-D(E_{f\downarrow})}{D(E_{f\uparrow})+D(E_{f\downarrow})}
\end{equation}
%++++++++++++
where $D(E_{f\uparrow})$ and $D(E_{f\downarrow})$ represent the density of states (DOS) of majority spin and minority spin at the Fermi level, respectively.\\
%+++++++++++++++++++++++++++++++++++++++++++++++++++++++++++++++++++++++++
%+++++++++++++++++++++++++++++++++++++++++++++++++++++++++++++++++++++++++
\indent To have a better understand of the magnetic coupling of the TM-GY systems, we use the map of spin-polarized charge density (SCD) for all investigated systems. The SCD are defined as:
%+++++++++++++++
\begin{equation}
\rho(\vec r)=\rho\uparrow(\vec r)-\rho\downarrow(\vec r)
\end{equation}
%+++++++++++++++
where $\rho\uparrow(\vec r)$ and $\rho\downarrow(\vec r)$ are the spin up and the spin down charge density of TM-GY system, respectively.\\
%===============================================================
\begin{table*}
\caption{Calculated results of TM-chain-GY. h represents the
height between the adatom and GY sheet. $d_{AC}$ represents distance
between TM adatom and its nearest neighbor (NN) carbon atoms of GY
sheet. M is the total magnetic moment per unit cell. $\Delta E$ is
the energy difference defined as $\Delta E$=$E_{FM}$-$E_{AFM}$,
where $E_{FM}$ and $E_{AFM}$ are the total energies for
ferromagnetic and antiferromagnetic states in the same unit cell,
respectively. $E_a$ is the adsorption energy of TM-GY systems. $T_e$
is the charge transfer from the TM to GY derived from AIM. The
polarization (P) and the magnetic states (MS) are also included in
the table.}\label{tab1}
\begin{ruledtabular}
\begin{tabular}{ccccccccc}
\hline
 &h($\AA$)&$d_{AC}$($\AA$)&M($\mu_B$)&$\Delta E(meV)$&$E_{a}(eV)$&$T_e$&P&MS\\
\hline
V         &1.47 &2.29 &7.32  &-271 &-2.787 &1.16 &63\% &FM \\
Cr        &1.24 &2.26 &0     &42 &-1.190 &1.01 &0 \% &AFM\\
Mn        &0    &2.03 &7.63  &-193 &-0.907 &1.28 &12\% &FM \\
Fe        &0.74 &2.14 &5.67  &-11  &-2.914 &0.94 &14\% &FM \\
Co        &0.49 &2.12 &3.5   &-16  &-2.108 &0.76 &24\% &FM \\
\end{tabular}
\end{ruledtabular}
\end{table*}
%+++++++++++++++++++++++++++++++++++++++++++++++++++++++++++++++++++++++++
%+++++++++++++++
\section{Results and discussions}
\subsection{Adsorption Geometry and Stability of TM chain on the surface of GY}
\indent Before investigating the magnetic properties of TM-chains-GY systems, we study the stability of TM chains adsorbed GY systems. First of all , we considered the dimer configurations of TM (V, Cr, Mn, Fe and Co) on the surface of GY. The most stable adsorption site is the alkyne ring of GY for all single TM atom (V, Cr, Mn, Fe and Co)\cite{19}. Based on the most stable single TM atom adsorption configuration, we put the second TM on the surface. Two initial possible adsorption sites for the second TM are considered, as shown in Fig.\ref{fig1},(1) H1 site is neighboring hexagonal ring and (2) H2 site is neighboring alkyne ring around the first one. We find that CON2 (both TM atoms are trapped in alkyne ring by -C$\equiv$C-) is the most stable configuration, as shown in Fig.\ref{fig1}. The configurations of CON1 (one TM trapped in in alkyne ring and another adsorbed at the hollow site of hexagonal ring) are 1.097eV, 0.515eV, 0.573eV, 0.978eV, and 1.084eV per TM atom less stable than that of CON2 for V, Cr, Mn, Fe and Co, respectively. The results indicate that all the five TM dimers are chemisorbed on GY sheet with large adsorption energies (-2.780 eV , -1.256 eV , -0.942 eV , -3.096 eV and -2.200 eV per TM atom for V, Cr, Mn, Fe, and Co dimers, respectively). Moreover, the adsorption energy per TM adatom for V-dimer-GY and Fe-dimer-GY systems is more stable than corresponding single V and Fe adsorption systems(the adsorption energy for single TM adsorbed GY system is -1.672, -1.331, -1.433, -2.424, and -2.541eV for V, Cr, Mn, Fe, and Co, respectively). Such stability is derived from the existence of -C$\equiv$C- bond. The $p_x$ and $p_y$ of \emph{sp}-hybridized carbon atoms in the GY contribute to both $\sigma$-bond and $\pi$-bond, making it easily to couple with the TM adatom. Increasing the number of TM atoms, we find that the adsorption energy is similar with the dimer configuration. Therefore, it is expected that a regular TM nanowire with a single atom thickness is possible to be constructed in the surface of GY sheet. Our results demonstrate that GY sheet are promising template for TM nanowire in spintronics.\\
%+++++++++++++++++++++++++++++++++++++++++++++++++++++++++++++++++++++++++
\indent After achieving the TM dimer adsorption configuration, we investigate the TM chains adsorbed on surface of GY. In present work, we consider TM (V, Cr, Mn, Fe and Co) zigzag monoatomic chain as shown in Fig.\ref{fig1}. The adsorption energies and structure parameters of equilibrium TM-chain-GY are listed in Tab.\ref{tab1}. The considerable negative adsorption energies for all of the 3d TM-chain on GY indicate their chemisorption characteristics. The weakest chemisorption case is Mn-GY whose adsorption energy is -0.907 eV, whereas the strongest chemisorption one is the Fe-GY whose adsorption energy is -2.914 eV. The stability of TM-chain-GY systems is derived from the existence of the carbon atoms with the -C$\equiv$C- bond as reported in our previous study\cite{19}. The valence electrons of the TM atom couple to not only the $p_z$ but also the $p_{x-y}$ of carbon atoms of the TM-chain-GY systems. The results in Tab.\ref{tab1} indicate that as for the maximum (minimum) distance between the adatom of TM-chain and the nearest carbon atom ($d_{AC}$) is 2.29 $\AA$ (2.03 $\AA$) for V (Mn). All TM adatoms of TM-chain only slightly protrude out of the GY plane except Mn; the largest height (h) is 1.47 $\AA$ for V, whereas the minimum height is 0 $\AA$ for Mn. The result indicates that the size of Mn adatom matches the acetylene ring of GY very well.\\
%+++++++++++++++++++++++++++++++++++++++++++++++++++++++++++++++++++++++++
%===============================================================
\begin{table}
\caption{The calculated results of TM-dimer-GY. M is the total
magnetic moment per unit cell. $\Delta E$ is the energy difference
defined as $\Delta E$=$E_{FM}$-$E_{AFM}$, where $E_{FM}$ and
$E_{AFM}$ are the total energies for ferromagnetic and
antiferromagnetic states in the same unit cell, respectively.}\label{tabdimer}
\begin{ruledtabular}
\begin{tabular}{cccccc}
\hline
dimer&V&Cr&Mn&Fe&Co\\
\hline
M($\mu_B$)&6.12&0.0&7.55&5.50&2.92\\
$\Delta E$(meV)&-48&22&-122&-17&-8\\
\end{tabular}
\end{ruledtabular}
\end{table}
%===============================================================
\subsection{Magnetic Order and Electronic Structures of TM-chain-GY}
\begin{figure}
 \includegraphics[width=3.5in]{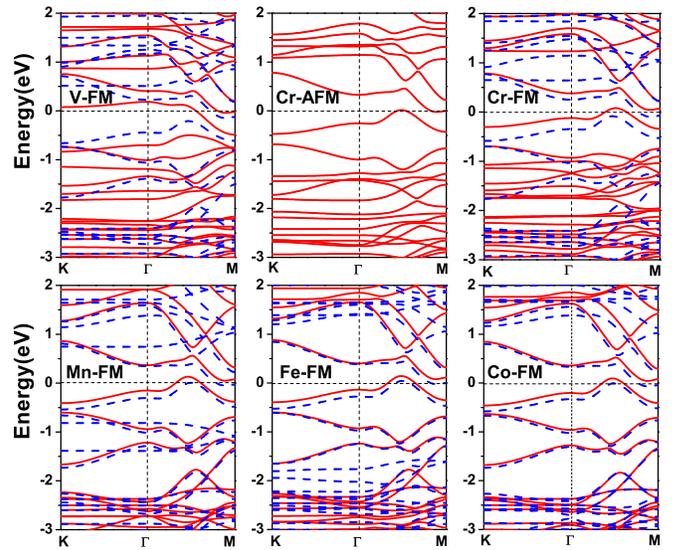}\\
 \caption{Band structures of V-chain-GY (a), Cr-chain-GY (c), Mn-chain-GY (d)
Fe -chain-GY(e), Co-chain-GY(f) under FM magnetic state, and Cr-chain-GY(b) under FM magnetic state, respectively. The solid (red) and dashed (blue) lines represent
the majority and minority spin channels, respectively. The Fermi
level is set to zero.}
 \label{fig2}
\end{figure}

\begin{figure}
 \includegraphics[width=3.5in]{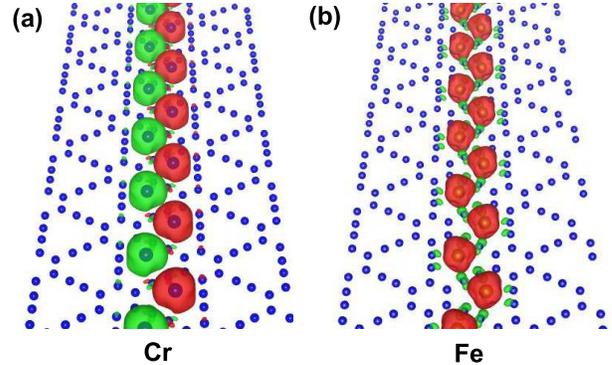}\\
 \caption{Spin-polarized charge density (SCD) distribution for Cr-chain-GY (a) and Fe-chain-GY (b). The dark (red) and light
(green) isosurfaces denote the spin-up and spin-down charge density, respectively. The isosurface of SCD is 0.004 e$/\AA^3$.}
 \label{fig3}
\end{figure}

\begin{figure}
 \includegraphics[width=3.5in]{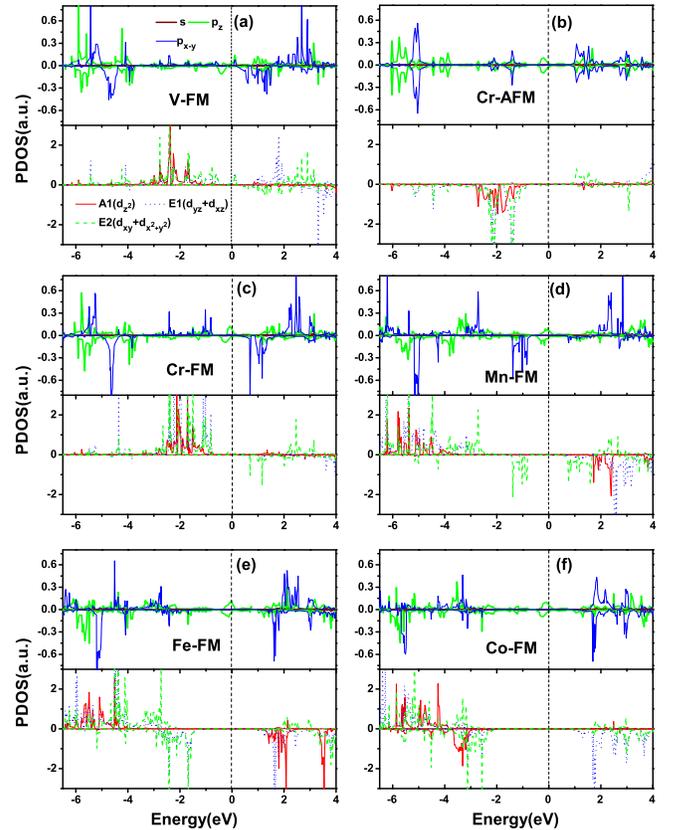}\\
 \caption{PDOS for V (a), Cr (c), Mn (d), Fe (e), and Co (f) under FM state and  Cr (b) under AFM states. The solid (red) and dash (blue) lines represent the majority and minority spin channels, respectively. The Fermi level is set to zero.}
 \label{fig4}
\end{figure}

\begin{figure}
 \includegraphics[width=3.5in]{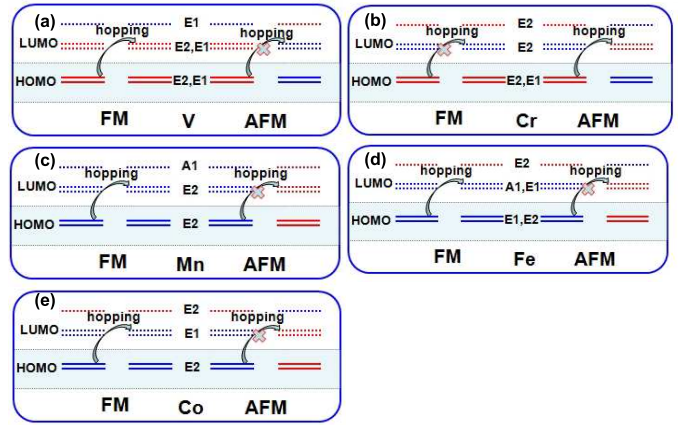}\\
 \caption{Illustration of energy level of $d$ orbital for V (a), Cr (b), Mn (c), Fe (d), and Co (e) and their virtual hopping mechanism.
 In each picture, left panel: virtual hopping under FM state; Right panel: virtual hopping under AFM
 state. Blue and red lines represent spin-up and spin-down level, respectively. Solid(dashed) line indicate occupied(unoccupied) states}
 \label{fig5}
\end{figure}

\indent On the basis of the most stable structures of TM-dimer-GY and TM-chain-GY, we systematically investigate their magnetic exchange coupling and electronic properties. Distinct spin polarization is found for the GY with adsorption of V, Cr, Mn, Fe, and Co dimer and chain. We consider only a collinear ferromagnetic (FM) and antiferromagnetic (AFM) coupling for each TM-dimer-GY and TM-chain-GY configuration at present. The exchange energy ($\Delta E$=$E_{FM}$-$E_{AFM}$) between the FM and AFM states per unit cell is calculated and shown in Tab.\ref{tabdimer} and Tab.\ref{tab1} for TM-dimer-GY and TM-chain-GY, respectively. Positive (negative) exchange energy indicates that the ground state of the system is AFM (FM). For the case of TM-dimer-GY, the two TM adatoms for V, Mn, Fe, and Co show ferromagnetic coupling characteristics, whereas Cr dimer shows antiferromagnetic coupling characteristics with zero magnetic moment. Moreover, for V and Mn dimer, the exchange energy is larger enough to stabilize their ferromagnetic nature at high temperature. Such magnetic coupling for TM-dimer-GY is derived form the mediation of the sp hybridized carbon atoms and their coupling effects are similar with the case of  TM-chain-GY will be discussed below. As for TM-chain-GY, from the band structures as shown in Fig.\ref{fig2}, one can see that the adsorption of TM chain on GY is an effective modulation approach to their electronic properties; and all TM-chain-GY systems show metallic characteristics. The TM-chain absorbed systems show similar band structures, which is derived from the quais-1D structural nature as the cases of carbon nanotubes\cite{46}, graphene nanoribbons\cite{47}, and Cr chain embedded graphene topological line defect\cite{48}. Although our previous investigation indicated that the GY sheet emerge magnetism by adsorption of single TM atom\cite{19}, the existence of magnetic order of  TM-chain-GY determined by the magnetic exchange coupling between TM atoms is deserved to be studied thoroughly. The results as listed in Tab.\ref{tab1} indicate that ferromagnetic state for V, Mn, Fe and Co chains is 271 meV, 193 meV, 11 meV and 16 meV energetically more favorable than that of antiferromagnetic state, respectively. Especially, the V-chain-GY and Mn-chain-GY show considerable exchange energy suggesting that V and Mn TM chain adsorbed on GY behave as stable ferromagnetism hopping to be high $T_c$ temperature magnetic materials. However, Cr chain on GY show antiferromagnetic coupling with zero net magnetic moment and its exchange energy is -42meV. The above results indicate that the TM chain on the surface of GY have robust FM and AFM order. To clearly show the magnetic coupling, we calculate the SCD and the results of typical ferromagnetic Fe-chain-GY and antiferromagnetic Cr-chain-GY are shown in Fig.\ref{fig3}. The results show that the magnetic moment of TM-chain-GY is mainly contributed by the TM adatom. Moreover, the SCD results indicate that the spin polarized charge density mainly localizes within the region of TM chain suggesting the coupling between neighbor chains is weak, which is crucial for achievement of a perfect spin-polarized current with TM chain in practical spintronic applications.\\
%+++++++++++++++++++++++++++++++++++++++++++++++++++++++++++++++++++++++++
%+++++++++++++++++++++++++++++++++++++++++++++++++++++++++++++++++++++++++
%+++++++++++++++++++++++++++++++++++++++++++++++++++++++++++++++++++++++++
\indent To understand the magnetic exchange coupling mechanism of TM-chain-GY, we show the partial density of states (PDOS) of TM and their nearest neighbor (NN) carbon atom in Fig.\ref{fig4}. We only show the PDOS of one of the NN \emph{sp}-hybridized carbon atoms of the TM adatom in the figure because the four NN carbon atoms of each TM adatom show similar characteristics. Here we take Fe-chains-GY as example to elucidate the relationship between the coupling and the magnetic properties of TM-chains-GY. The PDOS of Fe adatom and its NN carbon atoms for Fe-chain-GY are shown in Fig.\ref{fig4}(e). For the majority spin, the Fe \emph{d} orbital with A1($d_{z^2}$), E1 ($d_{yz}$,$d_{xz}$) and E2 ($d_{xy}$,$d_{x^2+y^2}$) symmetry strongly couples with the in-plane $p_{x-y}$ and out-plane $p_z$ of its NN C atoms in the energy ranges of -6.4eV to -2.1 eV. For the minority spin, one can see that there is considerable overlap for the A1, E1 and E2 states of Fe atom and $p_{x-y}$ and $p_z$ states of NN C atoms within -2.8 eV to -1.2 eV energy window. Because of the interaction of $p_z-d$ and $p_{x-y}-d$ around the Fermi level, the majority-spin \emph{p} band is shifted to higher energy, while the minority-spin \emph{p} band is shifted to lower energy. As shown in Fig.\ref{fig2}, bands around the Fermi level mainly derived from the $p_z$ state of the NN carbon atom, where the $p_z$ split to majority and minority band; and the majority bands shift higher energy, whereas the minority bands shift lower energy due to the effective \emph{pd} hybridization. The magnetic exchange coupling in such system follows the like-Zener's $p_z$-d exchange mechanism\cite{49,50} as that presents in diluted magnetic semiconductor\cite{51}. However, because of the \emph{d} states couples with not only the out-plane $p_z$ but also the in-plane $p_{x-y}$, the \emph{pd} hybridization shows the coexistence of out-of plane $p_z$-d and in-plane $p_{x-y}$-d exchange. The results in Fig\ref{fig4} clearly show that the $p_{x-y}$ states from \emph{sp} and $d$ states of TM overlap each other below the Fermi level indicating the key role of  $p_{x-y}$ states in the \emph{pd} hybridization process. Similar coupling mechanism can be find in the V-, Cr-, Mn- and Co-chain-GY systems. To analyze the magnetic coupling behaviour of TM-chain-GY, based on the results of PDOS, we illustrate the energy level of $d$ orbitals of neighboring TM atoms under both FM and AFM states as shown in Fig.\ref{fig5}. For example, the PDOS of V-chain-GY as shown in Fig.\ref{fig4}(a) indicates that the highest occupied (HOMO) and the lowest unoccupied (LUMO) \emph{d} orbital are both in spin up channel under FM state; whereas under AFM state, the HOMO is spin-up E1 and E2 states, and the LUMO is spin-down E1 and E2 states. Therefore, the favorable magnetic state of the V-chain-GY is FM due to virtual hopping between the frontier orbital and Hund's rules. The systems Mn-chain-GY, Fe-chain-GY, and Co-chain-GY have the similar virtual hopping characteristics; all of them show FM ground state. On the contrary, the virtual hopping for Cr-chain-GY, as shown in Fig.\ref{fig5}(b), will lead the system to AFM order. The results of  virtual hopping of the energy levels between neighboring TM atoms are agree well with the above results.\\
%+++++++++++++++++++++++++++++++++++++++++++++++++++++++++++++++++++++++++
\indent As mentioned above, TM-chain-GY systmes show strong coupling between TM-$d$ states and both out-plane $p_z$ and in-plane $p_{x-y}$ states of carbon atoms with $sp$ hybridization. Such coupling will induce electron exchange between TM and GY. To evaluate the charge transfer between the TM adatom and the GY sheet, the atomic basin charge based on the atoms in molecular (AIM) theory\cite{45} is adopted. By comparing the valence electrons in TM atomic basin of TM-chain-GY systems and those of corresponding TM free standing state, the charge transfer between TM and GY can be qualitatively determined. The electron transfer from TM adatom to GY sheet ($T_e$) as listed in Tab.\ref{tab1} indicates that there are considerable electron exchange from TM to GY sheet. Such electron exchange derives from the strong coupling between the \emph{d} states and \emph{p} states of the NN carbon, which is determined the magnetic states and magnetic moments of TM chain.\\
%+++++++++++++++++++++++++++++++++++++++++++++++++++++++++++++++++++++++++
%+++++++++++++++++++++++++++++++++++++++++++++++++++++++++++++++++++++++++
%===============================================================
%===============================================================
\begin{table*}
\caption{Spin magnetic moment($m_s$) and orbit magnetic moment($m_o$) per TM atom,
and MAE[x-z] and MAE[y-z] per unit cell of TM-chain-GY. All
the spin magnetic moments are presented for the easy axis of
magnetization.}\label{tab3}
\begin{ruledtabular}
\begin{tabular}{cccccccc}
\hline
 &MAE[x-z](meV)&MAE[y-z](meV)&$m_s$($\mu_B$)&$m_o$(easy)($\mu_B$)&$m_o$(hard)($\mu_B$)&Easy axis($\theta$,$\phi$)&Hard axis($\theta$,$\phi$) \\
\hline
V         &0.294  &0.816    &3.37   &0.094        &0.047       &(0,0) &(90,90) \\
Cr(AFM)   &0.612  &0.607    &4.33   &0.012        &0.007       &(0,0) &(90,0)  \\
Cr(FM)    &0.541  &0.605    &4.33   &0.019        &0.005       &(0,0) &(90,90) \\
Mn        &1.065  &1.054    &3.98   &0.040        &0.005       &(0,0) &(90,0)  \\
Fe        &2.232  &2.422    &2.95   &0.112        &0.089       &(0,0) &(90,90) \\
Co        &-2.076 &-1.750   &1.99   &0.230        &0.105       &(90,0) &(0,0)  \\
\end{tabular}
\end{ruledtabular}
\end{table*}
%===============================================================
\begin{figure}
 \includegraphics[width=3.5in]{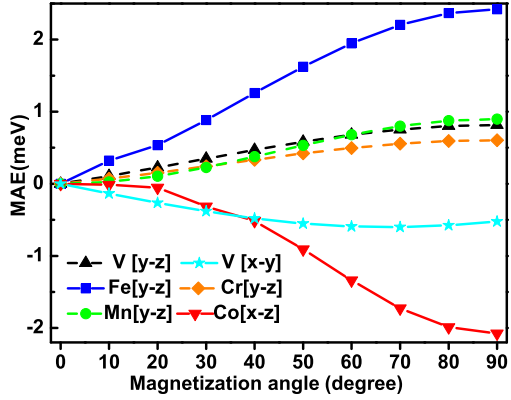}\\
 \caption{MAE as a function of magnetization angle, from x to z axis, y to z axis and x to y axis for hard axis to easy axis of TM-chain-GY, where the Cr-chain only show the FM states.}
 \label{fig6}
\end{figure}
%===============================================================
\subsection{Magnetic anisotropy of TM-chain-GY}
\indent Due to the reduced dimensions for TM chain, magnetic anisotropy can be enhanced and show unique properties. In previous study, it was indicated that 1D (chain or wire) nanostructure on metal surfaces shown lager MAE than 3D bulk material.\cite{52,59,60} Therefore, it is interesting to investigate magnetic anisotropy of 1D TM nanowire on the surface of GY. One of the most interesting issues is that : Is the easy axis parallel or perpendicular to the GY plane?\\
%===============================================================
\indent To this end, we perform noncollinear magnetic calculation using the GGA(PBE) exchange-correction functional for TM-chain-GY. It well known that the MAE is induced by spin-orbit coupling (SOC) effect. As for MAE calculations, the SOC effect are included. The spin magnetic moment under easy axis of magnetization, orbital magnetic moment under both easy and hard axis of magnetization for TM chain are calculated and the results are listed in Tab.\ref{tab3}. For easy axis of magnetization, the sequence of orbital magnetic moment of TM adatom are Co$>$Fe$>$V$>$Mn$>$Cr. Especially, Fe and Co shown considerable orbital magnetic moment with 0.112 $\mu_B$ and the 0.230 $\mu_B$. Moreover, the orbital magnetic moment of Co-chain-GY is similar to that of Co chain on the Pt(111) surface\cite{52} and Co dimer adsorbed on graphene\cite{53,54}. To characterize the MAE of TM-chain-GY systems, we perform calculations of the total energy while rotating the magnetization orientation (the magnetization orientation are defined by azimuth angle $\theta$ and polar angle $\phi$ as shown in Fig.\ref{fig1}, $\theta$ and $\phi$ use the z axis and x axis, respectively, as reference Cartesian axes): (i) from \emph{x} axis
(90,90) to \emph{z} axis(0,90) within xz plane; (ii)  from \emph{y} axis (90,0) to \emph{z} axis(90,90) within yz plane; (iii) from \emph{x} axis (90,90) to \emph{z} axis(90,0) within xy plane by 10 degree intervals with the definition of MAE per unit cell as follows:
%+++++++++++++++
\begin{equation}
MAE_{xz}=E(90,90)-E(\theta,90),  \   \   \theta\in(90,0)
\end{equation}
%+++++++++++++++
\begin{equation}
MAE_{yz}=E(90,0)-E(\theta,0), \   \   \theta\in(90,0)
\end{equation}
%+++++++++++++++
%+++++++++++++++
\begin{equation}
MAE_{xy}=E(90,90)-E(90,\phi), \    \   \phi\in(90,0)
\end{equation}
%+++++++++++++++
where $\theta$ and $\phi$ are the the azimuth and polar angles defined the magnetization orientation, as shown in Figure \ref{fig1}, the sketch of Cartesian coordination is also included in the figure. E($\theta$, $\phi$) is the total energy per unit cell when the magnetization orientation is ($\theta$, $\phi$). Figure \ref{fig6} summarizes the $MAE_{xz}$ $MAE_{yz}$ and $MAE_{xy}$ as a function of rotation angle for all TM-chain-GY systems. The results indicate that Fe-chain-GY and Co-chain-GY show relatively larger MAE as 2.422 meV and -2.076 meV, respectively. Fe-chain-GY and Co-chain-GY show easy-axis anisotropy and easy-plane anisotropy,respectively. V-, Cr-, and Mn-chain-GY systems show easy-axis anisotropy and their MAEs are 0.816 meV, 0.605 meV and 1.054 meV, respectively. Our results also indicate that when the quantization axis lies in the \emph{xy} plane the MAE is smaller except for V-chain-GY which exhibits 0.522 meV in-plane anisotropy as shown in Fig.\ref{fig6}. \\
%===============================================================
\indent The origin of magnetic anisotropy of TM-chain-GY is the SOC effect of the systems. The spin-conserving term of the SOC Hamilton $\hat{H}_{SO}=\lambda
\hat{L}\cdot \hat{S}$ can be defined as:\cite{55,56}
%+++++++++++++++
\begin{equation}
\hat{H}_{SO}^{sc}=\lambda
\hat{S}_n(\hat{L}_zcos\theta+\frac{1}{2}\hat{L}_+e^{-i\phi}sin\theta
 + \frac{1}{2}\hat{L}_-e^{i\phi}sin\theta) \label{eq7}
\end{equation}
%+++++++++++++++
Where the term $\lambda$ is the SCO constant, $\theta$ is the azimuthal angles (as shown in Fig.\ref{fig1}) of the magnetisation direction with the coordinate of TM-GY. By treating the SOC Hamilton as the second perturbations, the occupied( $\psiup_o$) and unoccupied($\psiup_u$) states can interact via the matrix element $<\psiup_u|\hat{H}_{SO}^{sc}|\psiup_o>$, the associated energy  lowering can be expressed in simplified form as:\cite{55,56}
%+++++++++++++++
\begin{equation}
 \Delta E_{SOC} \propto \sum_{\psiup_u,\psiup_o} \frac{<\psiup_u|\hat{H}_{SO}^{sc}|\psiup_o>^2}{E_\emph{u}-E_\emph{o}} \label{eq8}
\end{equation}
%+++++++++++++++
The interaction between occupied and unoccupied states determines the preferred spin orientations. The $d$ states with the same $|m|$ value interact through the operator $\hat{L}_z$ give nonzero matrix element, the different $|m|$ values interact through the ladder operator $\hat{L}_+$ and $\hat{L}_-$.\cite{55,56}\\
%===============================================================
\begin{figure}
\includegraphics[width=3.5in]{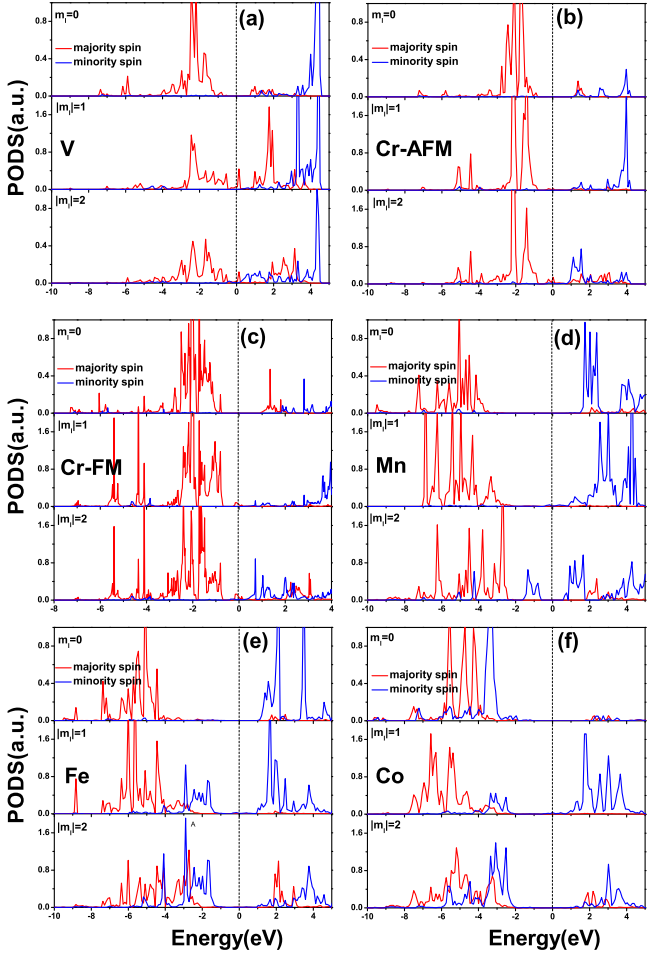}\\
\caption{Majority spin (red curves) and minority spin (blue curves) $m_l$-resolved PDOS for V (a), Cr (c), Mn (d), Fe (e) and  Co
(f) under FM state and Cr (b) under AFM state, respectively. The Fermi level is set to zero.}
 \label{fig7}
\end{figure}
%++++++++++++++++++++++++++++++++++++++++++++++++++++++++++++++++
\indent For V chain adsorbed on GY, the system exhibit easy-axis anisotropy. The results as shown in Fig.\ref{fig7}(a)  indicate that both HOMO and LUMO of V-chain-GY belong to the same spin channel (majority spin), and they are mainly derived from the $|m|$=1($d_{yz},d_{xz}$) and $|m|$=2($d_{x^2+y^2},d_{xy}$) states. Therefore, the energy lowering associated with the HOMO-LUMO interaction induced by SCO depends on the matrix element:
%+++++++++++++++
\begin{equation}
<d_{yz},d_{xz}|\hat{H}_{SO}^{sc}|d_{yz},d_{xz}> ,  \   \
<d_{x^2+y^2},d_{xy}|\hat{H}_{SO}^{sc}|d_{x^2+y^2},d_{xy}>
\end{equation}
%+++++++++++++++
Using the Eq.\ref{eq7}, we can easily get the nonzero matrix element depending on the magnetization direction as:
%+++++++++++++++
\begin{equation}
<d_{yz}|\hat{H}_{SO}^{sc}|d_{yz}> \propto
<Y_1^0|\hat{L}_z|Y_1^0>cos\theta
\end{equation}
%+++++++++++++++
Therefore, when $\theta$=$0^o$ (easy-axis anisotropy), the energy lowering is maximum, and the $d$ states of V with $|m|$=1 interact each other through the $\hat{L}_z$ operator. This is in agreement with the easy-axis anisotropy for V-chain-GY system. As for Cr-chain-GY, HOMO and LUMO state for both AFM and FM coupling are mainly derived from the $|m_l|$=2 states as shown in Fig.\ref{fig7}(b) and (c). Based on Eq.\ref{eq7},\ref{eq8}, one can see that the operator $\hat{L}_z$ couples the occupied and unoccupied states of $d_{x^2+y^2}$ and $d_{xy}$. Therefore, the coupling lowering the energy of  perpendicular magnetization direction, the systems show easy-axis anisotropy. From Fig. \ref{fig7}(d), one can see that the interaction between the HOMO and LUMO is mainly contributed by the $d_{x^2+y^2}$ and $d_{xy}$ states of minority spin for Mn-chain-GY system. The interaction between $d_{x^2+y^2}$ and $d_{xy}$ through the $\hat{L}_z$ leads to the easy-axis anisotropy. The $|m_l|$-resolved PDOS of Fe-chain-GY system as shown in Fig.\ref{fig7}(e) indicates that the HOMO is mainly derived from the $m_l$=1 and $|m_l|$=2 states, whereas the LUMO is mainly contributed by $|m_l|$=1 states. The SOC effect induces the interaction between $|m_l|$=1 of HOMO and LUMO by $\hat{L}_z$, which makes system show easy-axis anisotropy. However, the interaction between $|m_l|$=2 of HOMO and $|m_l|$=1 of LUMO by the ladder operator $\hat{L}_+$ is favorable for easy-plane anisotropy. The easy-axis anisotropy for Fe-chain-GY is derived from the competition effect between the operator $\hat{L}_+$ and $\hat{L}_z$ . From the $|m_l|$-resolved PDOS (Fig.\ref{fig7}(f)) of Co-chain-GY system one can see that the HOMO and LUMO are mainly contributed by $|m_l|$=2 and $|m_l|$=1 states. These states interact through the ladder operator $\hat{L}_+$ and $\hat{L}_z$. The energy-lowering associated with $\hat{H}_{SO}^{sc}$ approaches maximum when the $\theta$=90$^\circ$, i.e., the magnetization direction for Co-chain-GY is parallel to plane, indicating easy-plane anisotropy.\\
%+++++++++++++++
\section{Conclusion}
\indent In summary, we have performed DFT+U calculations on the magnetic exchange coupling and anisotropy of a quasi 1D 3d-TM zigzag nanowire adsorbed on GY sheet. TM intra-chain magnetic coupling mediated by \emph{sp} hybridized carbon atoms gives rise to robust long range ferromagnetic order except for Cr with anti-ferromagnetic order. The magnetic exchange interaction between TM adatom follows like-Zener's $p_z$-d exchange mechanism due to the coexistence of out-of plane $p_z$-d and in-plane $p_{xy}$-d exchange. We find that the Fe and Co chains show considerable magnetic anisotropy energy (MAE) and orbital magnetic moment. V, Cr, Mn and Fe chains show easy-axis anisotropy, whereas Co chian shows easy-plane anisotropy. Moreover, the in-plane anisotropy energy is close to zero except for V chain. \\
%+++++++++++++++++++++++++++++++++++++++++++++++++++++++++++++++++++++
\begin{acknowledgements}
This work is supported by the Program for New Century Excellent Talents in University (Grant No. NCET-10-0169), the Scientific Research Fund of Hunan Provincial Education Department (Grant No. 10K065), the National Natural Science Foundation of China (Grant Nos. 10874143,11274262, 11274029), the Hunan Provincial Innovation Foundation for Postgraduate (Grant No. CX2012B273).\\
\end{acknowledgements}
%\subsection{References}

\end{document}